\documentclass[]{aa}
\usepackage{txfonts}
\usepackage{graphicx}
\usepackage{natbib}
\bibpunct{(}{)}{;}{a}{}{,}

\newcommand{\edit}[1]{}
\newcommand{\new}[1]{{#1}}
\newcommand{\myremark}[1]{}

\def\eq{\begin{equation}}
\def\qe{\end{equation}}
\def\eqa{\begin{eqnarray}}
\def\aqe{\end{eqnarray}}

\newcommand{\refeq}[1]{Eq.~(\ref{#1})}                 
\newcommand{\refeqb}[1]{(Eq.~\ref{#1})}                 
\newcommand{\refsec}[1]{Sect.~\ref{#1}}                
\newcommand{\reffig}[1]{Fig.~\ref{#1}}                 

\newcommand{\CE}{C/E}

\newcommand{\xtenton}[2]{\ensuremath{#1 \cdot 10^{#2}}}      
\newcommand{\dfrac}[2]{\displaystyle\frac{#1}{#2}}

\newcommand{\Acld}{\ensuremath{A_{\rm cld}}}
\newcommand{\dtsys}{\ensuremath{\Delta t_{\rm sys}}}
\newcommand{\esn}{\ensuremath{\epsilon_{\rm SN}}}            
\newcommand{\ngas}{\ensuremath{n_{\rm gas}}}                
\newcommand{\Mtcld}{\ensuremath{M_{\rm cld}}}           
\newcommand{\Mcld}{\ensuremath{m_{\rm cld}}}           
\newcommand{\Msfc}{\ensuremath{M_{\rm SF,lim}}}           
\newcommand{\Pism}{\ensuremath{P_{\rm gas}}}           
\newcommand{\Psun}{\ensuremath{P_{\sun}}}           
\newcommand{\Psflim}{\ensuremath{P_{\rm SF,lim}}}      
\newcommand{\Rcld}{\ensuremath{r_{\rm cld}}}           
\newcommand{\soft}{\ensuremath{{\epsilon}_{\rm s}}}          
\newcommand{\Tgas}{\ensuremath{T_{\rm gas}}}           
\newcommand{\tia}{\ensuremath{\tau_{\rm ia}}}           

\def\sun{\hbox{$\odot$}}                                     

\newcommand{\cmmthree}{\ensuremath{\,{\rm cm}^{-3}}}         
\newcommand{\erg}{\ensuremath{\,{\rm erg}}}                  
\newcommand{\gram}{\ensuremath{\,{\rm g}}}                   
\newcommand{\gyr}{\ensuremath{\,{\rm Gyr}}}                  
\newcommand{\gyrmone}{\ensuremath{\,{\rm Gyr}^{-1}}}         
\newcommand{\Kelvin}{\ensuremath{\,{\rm K}}}                 
\newcommand{\kms}{\ensuremath{\,{\rm km}\,{\rm s}^{-1}}}     
\newcommand{\kpc}{\ensuremath{\,{\rm kpc}}}                  
\newcommand{\msun}{\ensuremath{\,{\rm M}_{\sun}}}            
\newcommand{\myr}{\ensuremath{\,{\rm Myr}}}                  
\newcommand{\pctwo}{\ensuremath{\,{\rm pc}^2}}               
\newcommand{\pcmtwo}{\ensuremath{\,{\rm pc}^{-2}}}           
\newcommand{\pc}{\ensuremath{\,{\rm pc}}}                    
\newcommand{\secmone}{\ensuremath{\,{\rm s}^{-1}}}           
\newcommand{\yrmone}{\ensuremath{\,{\rm yr}^{-1}}}           

\defcitealias{KD95}{KD95}
\defcitealias{EE97}{EE97}

\begin{document}
   \title{Modelling Galaxies with a 3d Multi-Phase ISM}

   \subtitle{}

   \author{S.~Harfst\inst{1,2}
          \and
       Ch.~Theis\inst{1,3}
      \and
       G.~Hensler\inst{3}
          }

   \offprints{S.~Harfst \email{harfst@astrophysik.uni-kiel.de}}

   \institute{Institut f\"ur Theoretische Physik und Astrophysik,
              Universit\"at Kiel, 24098 Kiel, Germany
         \and Centre for Astrophysics and Supercomputing, Swinburne
              University, Hawthorn,
              Victoria 3122, Australia
         \and Institut f\"ur Astronomie, Universit\"at Wien,
              T\"urkenschanzstr. 17, 1180 Wien, Austria }


        \abstract{We present a new particle code for modelling the
        evolution of galaxies. The code is based on a multi-phase
        description for the interstellar medium (ISM). We included
        star formation (SF), stellar feedback by massive stars and
        planetary nebulae, phase transitions and interactions between
        gas clouds and ambient diffuse gas, namely condensation,
        evaporation, drag and energy dissipation. The latter is
        realised by radiative cooling and inelastic cloud-cloud
        collisions. We present new schemes for SF and stellar
        feedback. They include a consistent calculation of the star
        formation efficiency (SFE) based on ISM properties as well as
        a detailed redistribution of the feedback energy into the
        different ISM phases.

        As a first test example we show a model of the evolution of a
        present day Milky-Way-type galaxy. Though the model exhibits a
        quasi-stationary behaviour in global properties like mass
        fractions or surface densities, the evolution of the ISM is
        locally strongly variable depending on the local SF and
        stellar feedback. We start only with two distinct phases, but
        a three-phase ISM is formed soon consisting of cold molecular
        clouds, a warm gas disk and a hot gaseous halo. Hot gas is
        also found in bubbles in the disk accompanied by type II
        supernovae explosions. The volume filling factor of the
        hot gas in the disk is $\sim35\%$. The mass spectrum of the
        clouds follows a power-law with an index of $\alpha \approx
        -2$. The star formation rate (SFR) is $\sim 1.6 \msun\yrmone$
        on average decreasing slowly with time due to gas
        consumption. In order to maintain a constant SFR gas
        replenishment, e.g.\ by infall, of the order $1 \msun\yrmone$
        is required.  Our model is in fair agreement with Kennicutt's
        (1998) SF law including the cut-off at $\sim10\msun\pcmtwo$.
        Models with a constant SFE, i.e.\ no feedback on the SF, fail
        to reproduce Kennicutt's law.

        We have performed a parameter study varying the particle
        resolution, feedback energy, cloud radius, SF time scale and
        metallicity. In most these cases the evolution of the model
        galaxy was not significantly different to our reference
        model. Increasing the feedback energy by a factor of $4-5$
        lowers the SF rate by $\sim 0.5\msun\yrmone$ and decreasing
        the metallicity by a factor of $\sim100$ increases the mass
        fraction of the hot gas from about 10\% to 30\%.

    }

   \maketitle
%
%

\section{Introduction}

The treatment of the instellar medium (ISM) is a crucial part in
modelling galaxies due to the importance of dissipation, star
formation (SF) and feedback.  Single-phase TREE-SPH models suffer
from over-cooling \citep{PJF99,TTP00} because the multi-phase
structure of the ISM is not resolved. As a result baryonic matter
loses angular momentum to the dark matter (DM) halos and the disks
formed in these simulations are too concentrated. Also, SF is overly
efficient. On the other hand, a multi-phase ISM is a
pre-requisite for the classical grid-based chemo-dynamical (CD) models
\citep{BTH92,TBH92,SHT97,SG03} but they lack the geometrical
flexibility of a (3d-)particle code.

An approach to include a multi-phase treatment of the ISM in a
particle code has been presented by \citet{SC02}, who combined
Smoothed Particles Hydrodynamics (SPH) with a sticky particle method
\citep{C84,CG85,N88,PJK93,TH93}. In the multi-phase ISM code of 
\citet{SC02} sticky particles are used to describe a cold, dense cloudy 
medium while the diffuse ISM is modelled by SPH. Stars form from cold
gas and feedback processes include heating of cold gas by supernovae
and stellar mass loss (including metal enrichment). Cooling leads to a
phase transition from warm to cold gas. The main advantage of this
approach is the dynamical treatment of the cloud component: clouds are
assumed to move on ballistic orbits dissipating kinetic energy in
collisions.

The main purpose of this paper is to present a new extended TREE-SPH
code with a multi-phase description of the ISM adopted from CD models.
Similar to \citet{SC02}, this is achieved by a combination of a sticky
particle scheme and SPH. However, we use a different sticky particle
scheme based on an individual treatment of single clouds. Advantages
of this scheme are the build-up of a cloud mass spectrum, the
individual handling of cloud collisions and the localised treatment of
phase transitions between clouds and the ambient diffuse gas.
Furthermore, we apply a refined SF recipe suggested by \citet[][
hereafter EE97]{EE97}. This scheme considers star formation within
individual clouds including local feedback processes. As a result the
star formation efficiency is locally variable depending on the cloud
mass and the ambient ISM pressure.

A detailed description of the model including all the processes is
given in \refsec{harfst_sec_model}. Details on the numerical treatment
can be found in \refsec{harfst_sec_nt}. The code is applied to an
isolated Milky-Way-type galaxy (\refsec{sec_model}). The results
(including a short parameter study) are discussed in
\refsec{sec_disc}.  Conclusions and future applications of this code
are presented in \refsec{sec_conclusions}.

%
\section{The Model}
\label{harfst_sec_model}

\begin{figure}[t]
\begin{center}
\resizebox{.7\hsize}{!}{\includegraphics{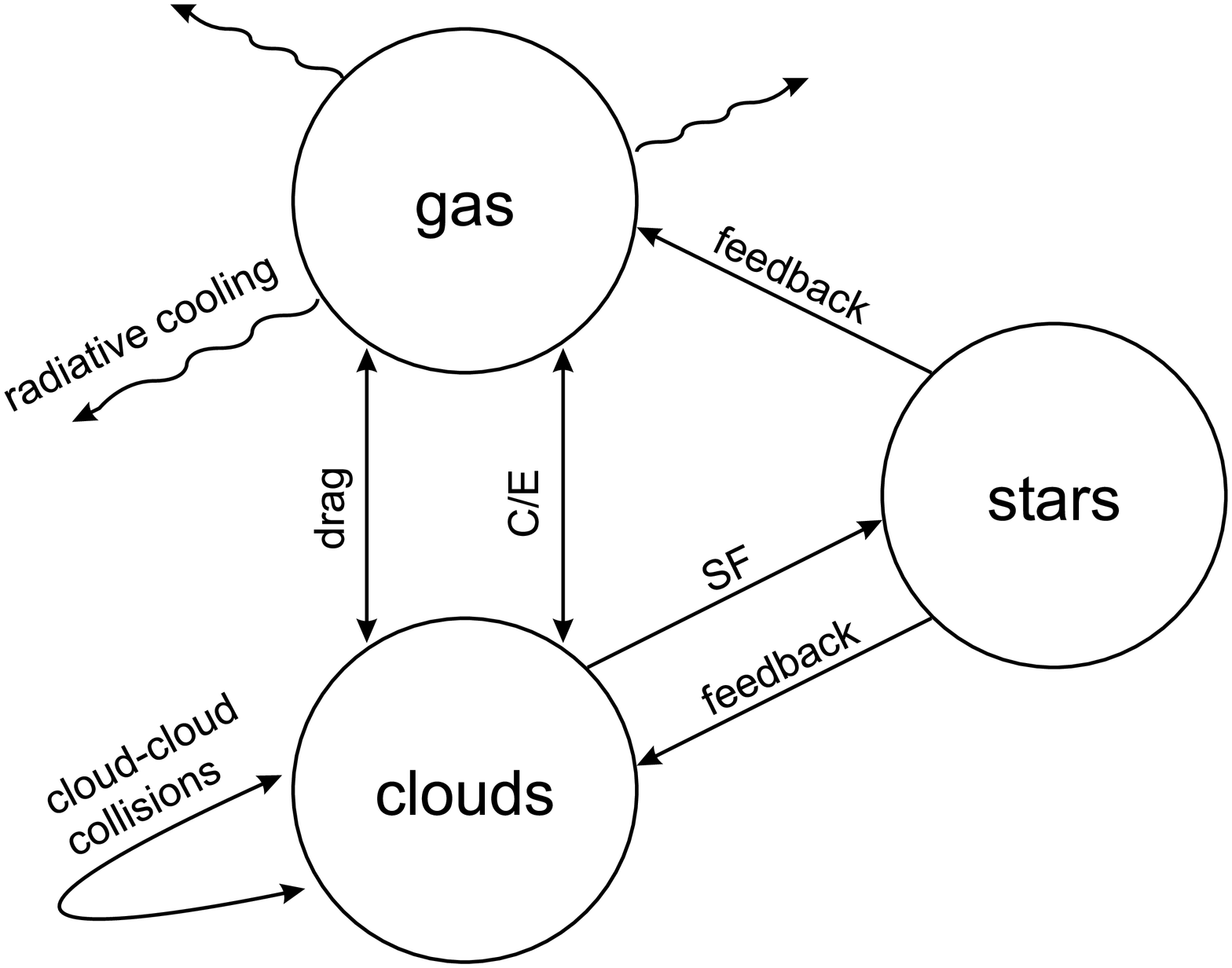}}
\end{center}
\caption{ The network of processes connecting the different
gaseous and stellar components. Clouds are the sites of SF.
Stars return mass and inject energy to the multi-phase ISM.
The two gas phases exchange mass and momentum by means of condensation
or evaporation and
drag. Dissipation of energy is due to radiative cooling and
cloud-cloud collisions.  } \label{fig_NT1}
\end{figure}

We use particles to describe all galactic components. Each
particle is assigned a special type representing either stars,
clouds, (diffuse) gas or DM. The different components are
connected by a network of processes including SF, stellar
feedback, cooling, cloud-cloud collisions, condensation,
evaporation and drag as shown schematically in \reffig{fig_NT1}.

\subsection{Stars}
\label{sec_stars}

Each star particle represents a cluster of stars formed at the same
time. The initial stellar mass distribution is given by the
multi-power law IMF of \citet{KTG93} with the lower and upper
mass limits of $0.1\msun$ to $100\msun$ respectively.

The stellar life times (depending on stellar mass and metallicity) are
approximated according to \citet{RVN96} to give reasonable fits to
the Padova evolutionary tracks \citep{ABB93,BFB93,BBC94} in the mass
range from $0.6\msun$ to $120\msun$ and for metallicities ${\rm Z} = 7
\cdot 10^{-5} - 0.03$.

We assume that stars with masses above $8\msun$ end their life as
type II supernovae (SNII) and stars from $1-8\msun$ end as planetary
nebulae (PNe). The mass fractions of these stars are $9\%$ and $34\%$
respectively. The life times of the high mass stars range from $3\myr$
to $39\myr$. The stellar feedback is described below.

\subsection{Clouds}
\label{sec_clouds}

The cloud particles describe a cold, clumpy phase of the ISM, i.e.\
the molecular clouds, which are also the sites of SF.  The radius
$\Rcld$ of an individual cloud is given by the following relation
based on observations \citep{RS88}:
\eq
r_{\rm cld} = \Acld \cdot \sqrt{\Mcld \over 10^6 \msun} \pc ,
\label{eq_mrrel}
\qe
\noindent
where the constant $\Acld$ is set to $50$.

The giant molecular clouds represented by cloud particles consist of
gas at different densities and temperatures. Since we do not resolve
this structure we do not define a temperature or pressure for
individual clouds. However, we assume that the clouds are influenced
by the pressure of the surrounding ISM (see
Sect.~\ref{sec_starformation}). Radiative cooling within clouds is not
taken into account.  However, we allow for dissipation of kinetic
energy in cloud-cloud collisions using the dissipation scheme of
\citet{TH93}.  Two clouds undergo a complete inelastic collision if
they physically collide within the next time step.  Additionally,
their orbital angular momentum must be less than the break-up spin of
the final cloud and the cloud mass must not exceed a critical mass
limit (here: $M_{\rm coll,lim} = \xtenton{1}{7}\msun$).  In case of a
sticky collision the two cloud particles are replaced by one particle
keeping the center-of-mass data of the former two clouds.

\subsection{Diffuse gas}

In an first approximation the diffuse gas can be described as
continuous self-gravitating fluid. We model this component using an
SPH approach. The gas is assumed to obey the ideal equation of state
with adiabatic index $\gamma=5/3$: 
\eq 
\Pism=(\gamma-1)\rho_{\rm gas} u,
\label{eq_eos}
\qe 
where $\Pism$ is the pressure, $\rho$ the density and $u$ the
energy per unit mass of the gas. The thermal energy evolution is
governed by cooling and heating processes. For the latter we only take
the most dominant source, the SNII, into account.  Dissipation is
performed by the metal-dependent cooling function from \citet{BH89}.

The diffuse gas can have temperatures from a few $10^3\Kelvin$ up to
several $10^6\Kelvin$. In the following we will also distinguish
between warm and hot gas meaning gas at temperatures below or
above $10^4\Kelvin$ respectively.

\subsection{Dark Matter}

In our code DM can be modelled either with particles (live halo) or as
a static potential.  In this work we focus on the late evolution
of an isolated galaxy where a static DM potential is a reasonable
approximation. We use the potential from \citet{KD95} which is a
flattened, isothermal potential with a core and a finite extent.

\subsection{Interactions between the ISM phases}

The ISM is treated as two dynamically independent gas phases. This
approach has been successfully applied in chemo-dynamical models
\citep{TBH92, SHT97}. The two gas phases are linked by two
processes: condensation/evaporation (\CE) and a drag force.  Following
\citet{CMO81} the mass exchange rate by condensation and evaporation
for an individual cloud particle is calculated from\footnote{We
changed the factor for the case $\sigma_{0}>1$ from
$\xtenton{-3.75}{4}$ to $\xtenton{-2.75}{4}$ in order to avoid a
discontinuity at $\sigma_{0}=1$.}
\eq
     \renewcommand{\arraystretch}{1.2}
     \begin{array}{ll}
     \dot{m}_{\rm cld} & = \left({\displaystyle\frac{\Tgas}{\Kelvin}}\right)^{2.5}
                   {\displaystyle\frac{\Rcld}{\pc}} \\
             & \cdot
        \left\{ \begin{array}{llllll}
                   \xtenton{-2.75}{4}\sigma_{0}^{-5/8}
                & {\rm if\ } &      &   & \sigma_{0} & > 1 \\
                   \xtenton{-2.75}{4}
                & {\rm if\ } & 0.03 & < & \sigma_{0} & \le 1\\
                   \xtenton{8.25}{2}\sigma_{0}^{-1}
                & {\rm if\ } &      &   & \sigma_{0} & \le 0.03
            \end{array}
        \right.\gram\secmone,
     \end{array}
\label{eq_ce}
\qe
where the parameter $\sigma_{0}$ is defined as
\eq
\sigma_{0} = \left({\Tgas \over \xtenton{1.54}{7}\Kelvin}\right)^2
             \left({\Rcld \over \pc}\right)^{-1}
             \left({\ngas \over \cmmthree}\right)^{-1} .
\label{eq_sigma0}
\qe

\noindent
$\Tgas$ and $\ngas$ are the local temperature and number density of
the diffuse gas. The clouds are subject to a drag force ${\bf F}_{\rm
D}$ due to ram pressure which is given by \citep{S72}
\eq
{\bf F}_{\rm D} = -C_{\rm D}\, \pi r_{\rm cld}^2\, \rho_{\rm gas}\,
v_{\rm rel}\, {\bf v}_{\rm rel},
\qe

\noindent
where $\rho_{\rm gas}$ is the local gas density and $v_{\rm rel}$ is
the velocity of the cloud relative to the gas. The constant $C_{\rm
D}$ is not well determined and it can be in the range $0.1-1$. We used
$C_{\rm D} = 0.1$.

\begin{figure}[t]
\resizebox{\hsize}{!}{\includegraphics{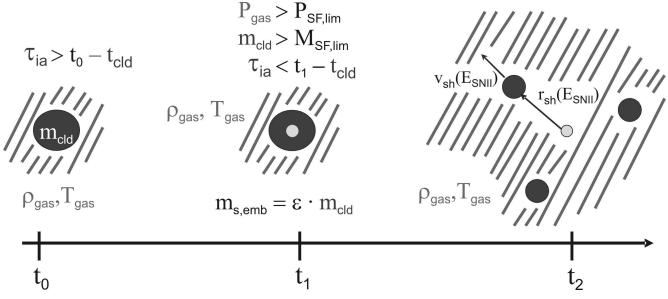}}
\caption{The diagram shows the SF scheme: ($t_0$) the cloud is
inactive (no SF); ($t_1$) an embedded star cluster has been formed
with a local SF efficiency; ($t_2$) the cloud is fragmented by SNII
energy input.  }
\label{fig_NT2}
\end{figure}

\subsection{Star Formation}
\label{sec_starformation}

The treatment of SF is usually based on the
Schmidt Law \citep{S59}
\eq
{\rm SFR/area} = c_n \cdot \Sigma_g^n,
\label{eq_schmidtlaw}
\qe

\noindent
with the gas surface density $\Sigma_g$ and $n \approx 1.5$. The
parameter $c_n$ combines a SF time scale and a SFE. A reasonable
estimate for the time scale is given by the dynamical time. The SFE on
the other hand cannot be determined from current theories of SF and
therefore remains a free parameter. A constant SFE also cannot
account for local feedback processes, by this prohibiting any
self-regulation of the SF. However, \citet{KTH95} have demonstrated
the importance of negative stellar feedback which leads to a
Schmidt-type strongly self-regulated SF.

Though the standard approach for SF works well for isolated galaxies,
it has difficulties in reproducing observations of interacting galaxies
\citep{MRB92,MBR93}.\citet{B04} suggested a second shock-induced SF
mode. However, this means another free parameter in the SF recipe.

Different to such an approach, we applied a single SF law, but with 
a temporally and spatially variable SFE suggested by 
\citetalias{EE97}.
They assign each molecular cloud an individual star formation efficiency
which is controlled by the ISM properties (mass of the molecular
cloud, pressure of ambient diffuse gas) and the stellar
feedback. According to \citetalias{EE97} the star formation 
is most efficient for massive clouds embedded in a high pressure ambient medium.  
A pre-requisite of using the EE97 SFE scheme is a multi-phase
treatment of the ISM as proposed in this paper.

The main concept of our SF implementation is that stars form in clouds
and clouds are destroyed by stellar feedback
and, thus, self-regulating the SF (\reffig{fig_NT2}). Each cloud is
assumed to form stars only after a time of inactivity $\tia$ (${\rm
  t}_0$ in \reffig{fig_NT2}), because not all gas in clouds is dense
or unstable enough for immediate SF. Once the SF process has started,
an embedded star cluster is formed (${\rm t}_1$) instantaneously
\footnote{We neglect the short time for the SF process itself.  This
  is motivated by recent observations suggesting that SF occurs on the
  local crossing time scale for sound waves in a cloud which is
  shorter than a few $10\myr$ \citep{E00}}.

Following \citetalias{EE97} we use a variable SF efficiency $\epsilon$
depending on local properties of the ISM.  A typical SFE (cloud mass
of $10^4-10^6\msun$ and pressure typical for solar vicinity) is
between 5\% and 10\% (see Fig.~4 in \citetalias{EE97}).  For details
of the implementation see App.~\ref{app_starformation}.

The only free parameter in our SF recipe is $\tia$.  It can be
interpreted as a global SF time scale: assuming a SFE of $\epsilon
\sim 0.1$, a total mass $\Mtcld$ in clouds of a few $10^9\msun$ and
$\tia$ of a few $100\myr$ , we get a global ${\rm SFR} \sim \epsilon
\Mtcld \tia^{-1}$ of $\sim 1\msun\yrmone$.

\subsection{Stellar Feedback}

Once the SFE has been determined the energy released by SNII in
each time step can be computed from the chosen IMF and stellar life
times (Sect.~\ref{sec_stars}). We assume
$\esn=\xtenton{2}{50}\erg$ for the energy released per SNII event.

This energy injection disrupts the cloud into smaller fragments (${\rm
t}_2$ in \reffig{fig_NT2}) leaving a new stellar particle. The time
for disruption $t_{\rm dis}$ ($= {\rm t}_2 - {\rm t}_1$) is determined
from the energy input: in our model the energy input of the SNII
drives an expanding shell into the cloud. The cloud mass is
accumulated in this shell. The expansion of the SN shell can be
calculated from a self-similar solution \citep{BBT95,TEP98}. Assuming
a spherical cloud with a $1/r$-density profile in agreement with the
mass-radius relation \refeqb{eq_mrrel} we get
\eq
\label{eq_rsh}
    r_{\rm sh}(t) = 0.918\left( {\dot{E} \over \rho_1} \right)^{0.25}
                       \cdot t^{0.75},
\qe
\eq
\label{eq_vsh}
    v_{\rm sh}(t) = 0.689\left( {\dot{E} \over \rho_1} \right)^{0.25}
                       \cdot t^{-0.25},
\qe
\noindent
where $\dot{E}$ is the energy injection rate and
$\rho_1 = \Mcld/\Rcld^2$.

It is assumed that the cloud is disrupted when the shell radius
reaches the cloud radius. The fragments are then placed symmetrically
on the shell with velocities corresponding to the expansion velocity
at $t_{\rm dis}$. The energy not consumed by cloud disruption, i.e.\
the energy of SNII exploding after $t_{\rm dis}$, is returned as
thermal feedback to the hot gas phase.
Depending on cloud mass and SF efficiency the fragmentation uses about
0.1--5\% of the total SNII energy of which $\sim20\%$ end up as
kinetic energy of the fragments.

The mass returned by SNII is also added to the surrounding
gas. It is determined from the IMF and stellar life times assuming
that the mass ejected per SNII event $M_{\rm SNII,ej}$ is given by
\citep{RVN96} 
\eq
 M_{\rm SNII,ej}(m)  = 0.77m^{1.06}\msun,
\qe 
where $m$ is the mass of a star.

\noindent
Additionally to the SNII feedback, we also allow for feedback by
PNe. After each time step, the mass $M_{\rm PN}$ returned to
the ISM by PNe is calculated by assuming
that the PN mass ejected by a star of mass $m$ is given by \citep{W00}
\eq
 M_{\rm PN,ej}(m)  = 0.91m-0.45\msun.
\qe

Gradual metal enrichment due to stellar mass release and type I SNe
are not included so far. This is justified for the kind of models
presented here (s.\ Sect.~\ref{sec_setup}) as metal enrichment has
little effect as we show later (s.\ Sect.~\ref{ssec_zres}).

%
%

%

\section{Numerical Treatment}
\label{harfst_sec_nt}

\subsection{Stellar Dynamics}

In our code the gravitational force on a particle can be calculated as
a combination of the self-gravity of the N-body system and an external
force. External forces -- if applied -- usually mimic a static dark
matter halo. Self-gravity is calculated using a tree scheme. We have
included two different tree schemes: the widely used scheme of
\citet{BH86} and a new tree algorithm proposed by \citet{D02}.  An
advantage of the Dehnen-TREE is that it is momentum-conserving by
construction. Additionally, an error control is included and the CPU
scaling is basically linear with the number of particles.  In our
implementation the Dehnen-TREE is more than an order of magnitude
faster than the Barnes\&Hut-TREE. Hence, we applied here mainly the
Dehnen-TREE.  For easy comparison with other N-body solvers we
realised gravitational softening by a Plummer potential with a
softening length $\soft = 0.1 \kpc$.

\subsection{Gas Dynamics}

We use the SPH formalism to compute the hydrodynamics of the diffuse
gas and a sticky particle scheme for the clouds. Details of our
SPH implementation are described in App.\ \ref{app_sph}.

The dynamics of the clouds are described by the sticky particle scheme of
\citet{TH93}. According to them clouds move on ballistic orbits unless
they undergo inelastic collisions. The treatment of cloud-cloud
collisions is described in \refsec{sec_clouds}.

\subsection{Time Integration}

The time integration of the system is done with a leap-frog
scheme. Individual time steps are used for each particle. The time
step for an SPH particle $p$ is chosen to satisfy a modified form of
the Courant condition (for details see App.\ \ref{app_sph}).
The time step of all other particles is based on dynamical criteria
calculated by
\eq
    \Delta t_p \le \eta_{\rm ts} \cdot \min\left( {\soft \over v_p}, \sqrt{\soft \over a_p } \right),
\qe

\noindent
where $a_p$ is the acceleration and $\soft$ the softening length of
particle $p$. The constant $\eta_{\rm ts}$ is of the order of unity (we set it
to 0.4).

The largest individual time step $\Delta t_p$ also defines the system
time step $\Delta t_{\rm sys}$ which is further limited not to exceed
$\sim1\myr$. Individual particle time steps are chosen to be a
power-of-two subdivision of $\Delta t_{\rm sys}$ \citep{Aarseth85}.

\subsection{Mass exchange and new particles}

Several processes included in our model lead to an exchange of
mass between particles or the creation or removal of
particles. These processes are introduced in a matter- and
momentum-conserving manner.

The mass exchange by condensation and evaporation is calculated
after each system time step $\dtsys$. The transferred mass
$\dot{m}_{\rm cld}\dtsys$ is added to the cloud. The same mass is
subtracted from the surrounding SPH particles taking into the
account the weighting from the kernel function. Additionally, the
velocity of particles gaining mass (i.e.\ clouds in case of
condensation and SPH particles for evaporation) is changed in
order to conserve momentum. The mass exchange from stars to SPH
particles (SNII) and clouds (PNe) is done in a similar way.

Whenever a SF process occurs, the star forming cloud is removed from
the simulation. Instead a new star particle of the mass
$\epsilon\Mcld$ is added at the same position with the same
velocity. Furthermore, the clouds formed in expanding shells are
created by adding four new cloud particles carrying the remaining
mass. Their positions and velocities are chosen according to
\refeq{eq_rsh} and (\ref{eq_vsh}) thus conserving mass and momentum.

\subsection{Mass limits for particles}

Our model allows particles to be created, merged and changed in
mass. A few constraints are used for the sake of numerical accuracy
and particle numbers.

In order to prevent the formation of too many low mass star particles
a lower mass limit $\Msfc = \xtenton{2.5}{4}\msun$ and a lower
pressure limit $\Psflim = 0.1\Psun$ for SF are introduced.

The maximum mass for clouds is given by $M_{\rm coll,lim}$
(\refsec{sec_clouds}) but a lower mass limit for clouds $M_{\rm
cld,lim}$ is also set to
\eq
   M_{\rm cld,lim} = 0.1 \Msfc.
\qe

\noindent
If the mass of a cloud becomes smaller than $M_{\rm cld,lim}$ the
particle is removed from the system and the mass is distributed evenly
among neighbouring cloud particles within $300\pc$.

SPH particles can change their mass due to condensation and
evaporation. In order to avoid SPH particles getting too massive an
upper mass limit is introduced. Whenever an SPH particle reaches this
limit it is split into four new particles. The new equal mass
particles have the same center of mass, velocity, angular momentum and
specific energy as the particle they replace. A lower mass limit
is also used where the mass of an SPH particle is distributed to the
neighbouring particles. The upper and lower mass limits are set to
$4\,m_{\rm SPH}$ and $0.1\,m_{\rm SPH}$ respectively, where $m_{\rm
SPH}$ is the SPH particle mass at the beginning of the simulation.

\subsection{Tests}

We have tested our code thoroughly. Energy and angular momentum
conservation is better than 1\% for a pure $N$-body simulation
following the evolution of a disk galaxy over many rotational periods.
We tested our SPH implementation with the standard Evrard-test
\citep{E88} and find that energy conservation is within 3\%. In a full
model, angular momentum is conserved within 3\%.

%
%

\section{Disk galaxy model}
\label{sec_model}

\subsection{Initial Conditions}
\label{sec_setup}

%
%
\begin{table}
   \centering
   \caption{Properties of the initial disk galaxy model}
   \label{tab_inimod}
   \setlength{\doublerulesep}{1pt}
   \renewcommand{\arraystretch}{1.2}
   \begin{tabular}{lccc}
   \hline\hline
   component          & mass$^1$ & radius$^2$ & No. of particles \\[0pt]
   \hline
   disk               & 0.34     &  18.0      & 20000            \\[-3pt]
   \hspace{5mm}stars  & 0.27     &            & 16000            \\[-3pt]
   \hspace{5mm}clouds & 0.07     &            & 4000             \\[0pt]
   bulge              & 0.17     &  4.0       & 10000            \\[0pt]
   DM halo            & 1.94     &  87.9      & -                \\[0pt]
   hot gas            & 0.01     &  25.       & 5000             \\[0pt]
   total              & 2.47     &            & 35000            \\[0pt]
   \hline\hline
  \multicolumn{3}{l}{$^1$ in units of $10^{11}\,{\rm M}_{\odot}$, $^2$ in kpc} \\
\end{tabular}
\end{table}
%

A galaxy similar to the present-day Milky Way was set up using the
galaxy building package described by \citet[][ hereafter KD95]{KD95}.
The models of \citetalias{KD95} consist of three components: the
baryonic disk and bulge, and a DM halo. A distribution function for
each component yields a unique density for a given potential. By
solving the Boltzmann equation for the combined system the galaxy can
be set up in a self-consistent way. By construction these models are
in virial equilibrium. \citetalias{KD95} have also demonstrated that
their models are stable over many rotation periods.

We chose model A of \citetalias{KD95} as a reference model
(Tab.~\ref{tab_inimod}). The scale length of the disk is $R_{\rm d} =
4\kpc$. The total mass of the galaxy is
$\xtenton{2.47}{11}\msun$. The mass and radius of the disk are set to
$\xtenton{0.34}{11}\msun$ and $18.0\kpc$. The corresponding parameters
for the bulge are $\xtenton{0.17}{11}\msun$ and $4.0\kpc$ and for the
DM halo $\xtenton{1.94}{11}\msun$ and $87.9\kpc$, respectively. The
particle numbers for disk and bulge, $N_{\rm d} = 20\,000$ and $N_{\rm
b} = 10\,000$, are chosen to have particles of nearly equal mass. The
DM halo is added as a static potential. The disk is initially in
rotational equilibrium with a small fraction of disordered motion
superimposed. This results in a Toomre parameter $Q\equiv
\kappa\sigma_R/(3.36G\Sigma)$ exceeding 1.9 all over the disk. Thus,
according to \citet{PPS97} the disk is stable against all axisymmetric
and non-axisymmetric perturbations.

A cloudy gas phase is added by treating every fifth particle from the
stellar disk as a cloud. The total mass in clouds is $\Mtcld \approx
6.9\cdot10^9\,{\rm M}_{\odot}$ and each cloud is randomly assigned a
time of inactivity $\tia$ between $0$ and $200\myr$. Initially, all
clouds have the same mass ($\sim\xtenton{1.7}{6}\msun$) but a mass
spectrum is built-up within the first Gyr of evolution.

Finally, a diffuse gas component with a total mass of ${\rm M}_{\rm
gas} = 1\cdot10^9\,{\rm M}_{\odot}$ is added. This is more difficult
because no simple equilibrium state exists for this gas phase: it is
not only affected by the gravitational potential but also by cooling
and heating processes. The latter varies locally and temporally so
that the diffuse gas can only reach a dynamical equilibrium as is also
revealed in high-resolution simulations of the ISM
\citep{AB04}. Testing different initial setups, starting far from
equilibrium (for the diffuse component only), turned out to be most
efficient.  Then, the diffuse gas phase relaxes on a few dynamical
timescales to a quasi-equilibrium state.  In detail, we started with a
slowly rotating homogeneous gaseous halo with a radius of $25\kpc$ and
a temperature of $\xtenton{5.4}{4}\Kelvin$ which relaxed to a
quasi-equilibrium within a few 100\,Myr.

Since we do not follow the chemical evolution of the galaxy, we apply
solar abundances where a metallicity is needed. In order to treat the
feedback of 'old' star particles each star particle is assigned an
age. For the bulge stars the age is $10\gyr$ and for disk stars
$0-10\gyr$.

\subsection{Evolution of the Reference Model}
\label{sec_refmod}

%
%
   \begin{figure} \resizebox{\hsize}{!}  {
   \includegraphics[width=\hsize,angle=90]{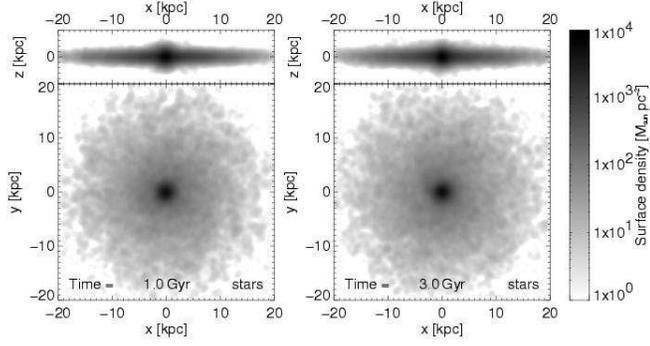}}
   \caption{The evolution of stellar surface density
   (at $1\gyr$ and $3\gyr$). The stellar disk becomes slightly thicker
   with time but is otherwise stable with weak transient spiral
   patterns.  The surface density plots (\reffig{fig_M1} and
   \ref{fig_M2}) were computed using the NEMO Stellar Dynamics Toolbox
   \citep{T95}. The particles were binned in a grid with a cell size
   of $40\pc$. The resulting map was then smoothed with a Gaussian
   kernel with ${\rm FWHM} = 1\kpc$.}
   \label{fig_M1} 
   \end{figure}
%

%
%
   \begin{figure}
   \resizebox{\hsize}{!}{
      \includegraphics[width=\hsize,angle=90]{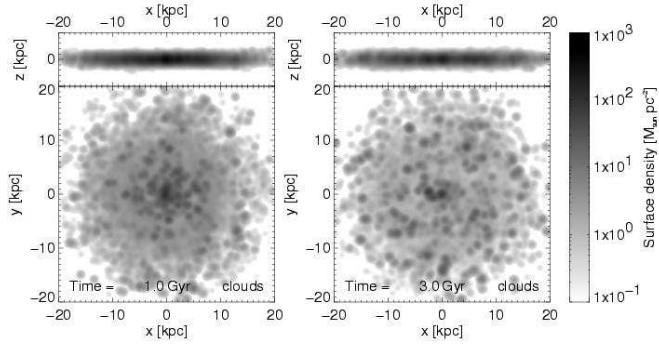}}
      \caption{ The evolution of cloud surface density (at
      $1\gyr$ and $3\gyr$). The cloud disk is stable but the
      overall surface density is slowly decreasing. For details on how
      the surface density was computed see \reffig{fig_M1}}
      \label{fig_M2}
\end{figure}
%

%
%
   \begin{figure}
      \resizebox{\hsize}{!}{
         \includegraphics[width=\hsize,angle=90]{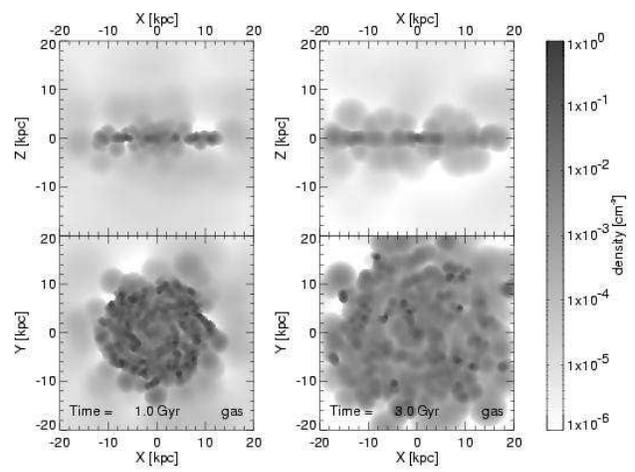}
	}
       \caption{ The of evolution of gas density (at $1\gyr$
       and $3\gyr$). In each panel the xy- (bottom),
       xz-projection (top) and density scale are shown.  Starting from
       a homogenous distribution the diffuse gas collapses and forms a
       thin gas disk within $1\gyr$. While the mass of the gaseous
       disk is nearly constant its radius is growing slowly. Densities
       in the disk are of the order of $1\cmmthree$. The halo has
       densities from $10^{-4}\cmmthree$ to
       $10^{-6}\cmmthree$. Densities were computed on a
       $100$x$100$-grid using the SPH formalism.}
       \label{fig_M3}
   \end{figure}
%

%
%
   \begin{figure}
   \resizebox{\hsize}{!}{
      \includegraphics[width=\hsize,angle=90]{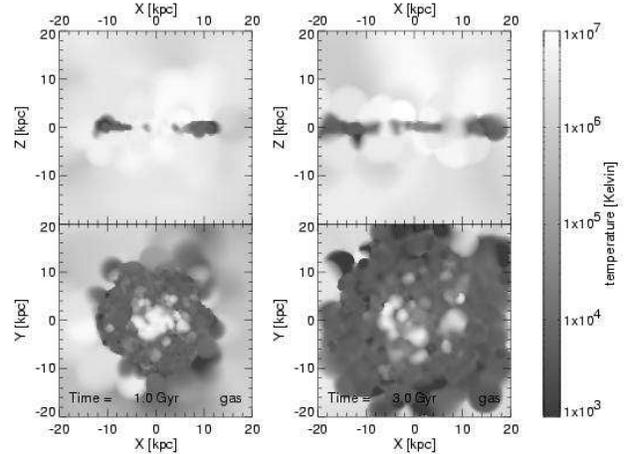} 
}
      \caption{ The evolution of gas temperature (at
        $1\gyr$ and $3\gyr$). In each panel the
        xy- (bottom), xz-projection (top) and temperature scale are
        shown. After the initial collapse the diffuse gas forms a disk
        at about $10^4\Kelvin$. The central part of the disk as well
        as small bubbles in the outer parts are heated by SNII to
        $10^6\Kelvin$. The halo gas it at $10^6-10^7\Kelvin$.}
         \label{fig_M4}
   \end{figure}
%

%
%
   \begin{figure} \resizebox{\hsize}{!}
   {\includegraphics[width=\hsize,angle=90]{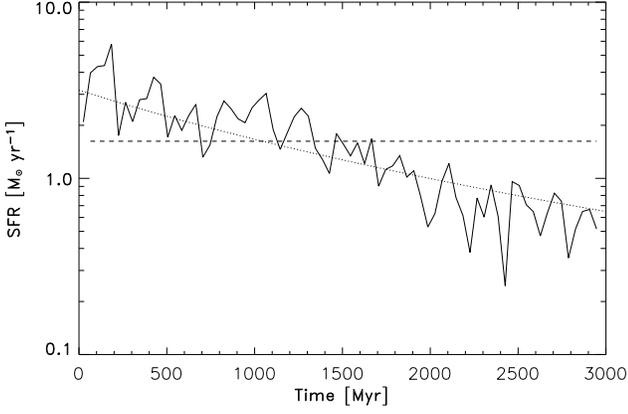}}
   \caption{ The evolution of the SFR with time (full line). The
   SFR decreases with time due to consumption of cloud mass. The
   average SFR (dashed line) is $\approx 1.6\msun\yrmone$. The
   dotted line is the result of a simple analytic model accounting for
   the depletion of the cloudy medium by SF.  } \label{fig_M6}
   \end{figure}
%

%
%
   \begin{figure} \resizebox{\hsize}{!}  {
   \includegraphics[width=\hsize,angle=90]{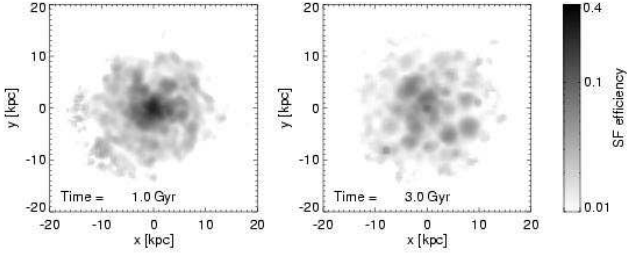} 
} 
\caption{ The evolution of SF efficiency . SFE
   is up to $0.4$ in the center of the galaxy. The average
   over the whole disk is about $0.06$.}
   \label{fig_M7}
\end{figure}
%

%
%
   \begin{figure} \resizebox{\hsize}{!}
   {\includegraphics[width=\hsize,angle=90]{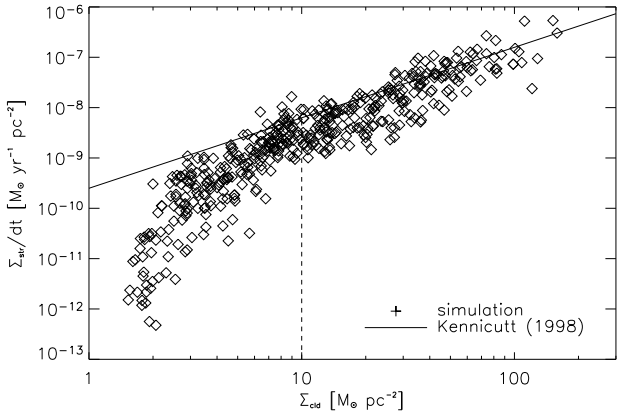}}
   \caption{ The open diamonds show the SFR per $\pctwo$ averaged over
   $100\myr$ as a function of cloud surface density. The full line is
   a recent determination of the Schmidt Law by \citet{K98} with a
   slope of $-1.4$. The vertical dashed line indicates a cut-off
   density for SF in disk galaxies also found by \citet{K98}.}
   \label{fig_M8} \end{figure}
%

%
%
   \begin{figure}
   \resizebox{\hsize}{!}
    {\includegraphics[width=\hsize,angle=90]{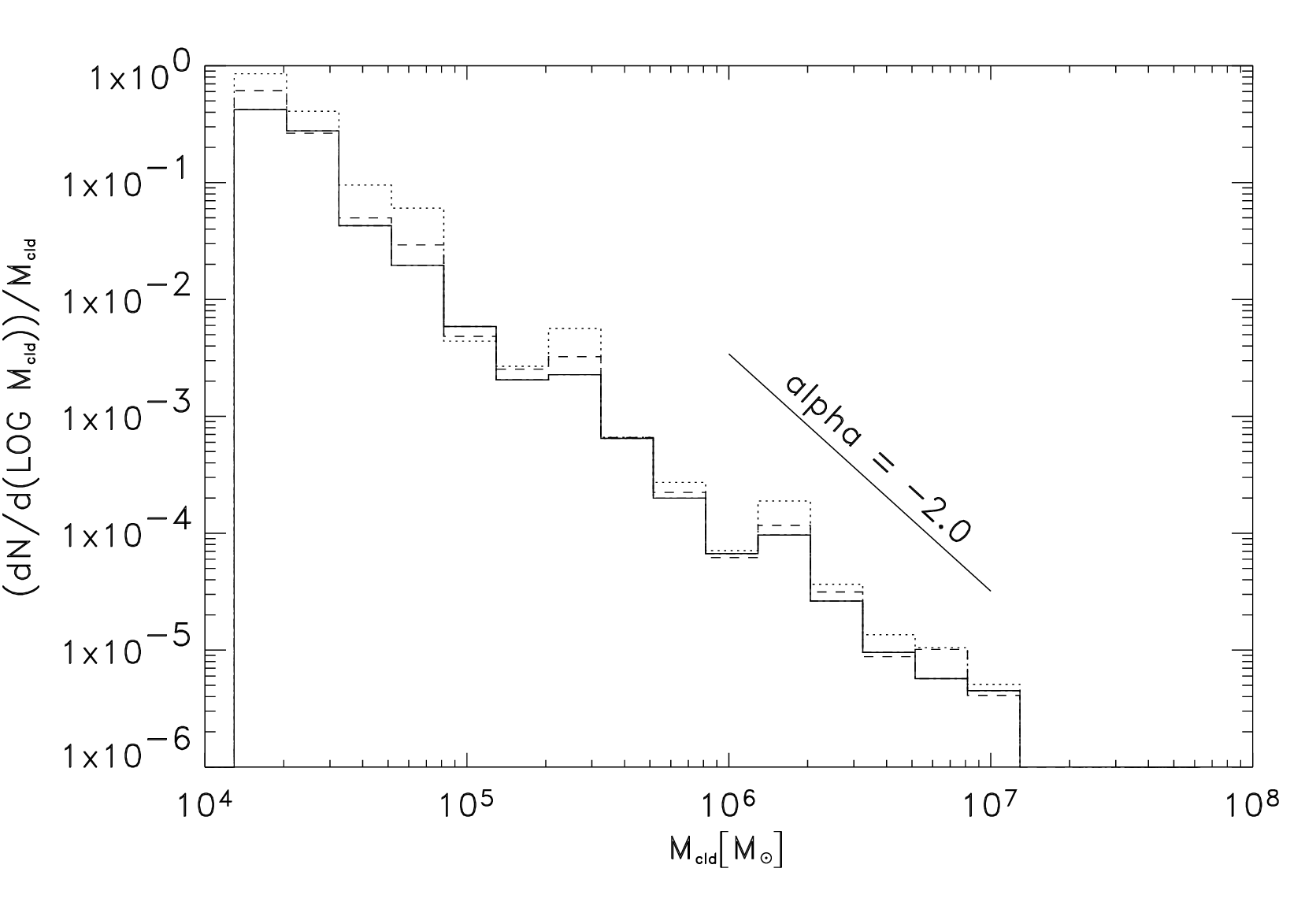}}
           \caption{ The cloud mass spectrum. The different lines correspond to
             the mass spectrum after $1\gyr$ (dotted), $2\gyr$
             (dashed) and $3\gyr$ (full) of evolution. The fitted
             power-law mass function is with $\alpha \approx -2.0$
             comparable to the observed mass spectrum.  }
         \label{fig_M11}
   \end{figure}
%

Starting from these initial conditions the galaxy model is evolved for
$3\gyr$ with all processes switched on. After an initial phase
($500\myr$) dominated by the collapse of the gaseous halo the galaxy
reaches a quasi-equilibrium defined by the interplay of the
gravitational potential, radiative cooling, heating by SNII and
condensation and evaporation. When the equilibrium is reached the
diffuse gas phase has formed a warm disk and a hot halo component.
Their mass ratio of about $7$ remains nearly constant for the rest of
the simulation. The equilibrium state reached is almost
independent of intial setup of the gas halo, i.e.\ the following
evolution is not influenced by our choice.

Snapshots of the evolution of the Milky Way model can be seen in
\reffig{fig_M1} to \ref{fig_M4}. The stellar surface density plots in
\reffig{fig_M1} demonstrate the stability of the disk during the
$3\gyr$ evolution.  Profiles of the surface density show that the
radial mass distribution does not change except for the central
$2\kpc$ where the surface density is increased by up to a factor of
$5$ caused by the dissipation-induced infall of clouds which
later form stars. Face-on views show weak transient spiral patterns,
e.g.\ at $t = 2\gyr$. Edge-on views reveal a thickening of the
disk. Throughout the disk the 95\%-mass height $z_{\rm 95,str}$
increases by 47\% from $0.75\kpc$ to about $1.1\kpc$. Only in the
central $2\kpc$ can no thickening be seen.

The evolution of the cloudy phase (\reffig{fig_M2}) is influenced by
dissipation and SF. The depletion of the cloudy medium by SF decreases
its mean surface density. Due to inelastic cloud-cloud collisions the
clouds also show a less smooth distribution. As a consequence a mass
flow towards the center increases its central surface density. The
radial profiles reveal that the surface density has decreased mostly
around $5\kpc$ while it it almost unchanged beyond $10\kpc$. The cloud
disk shows a thickening only outside of $11\kpc$ where the increase of
$z_{\rm 95,cld}$ is very similar to that of the stellar disk. Inside
that radius no thickening can be seen. In fact, due to dissipation
$z_{\rm 95,cld}$ is even reduced inside $5\kpc$ by a factor of 2--3.

The diffuse gas (\reffig{fig_M3}) starts from a homogeneous
distribution which collapses in less than $300\myr$ to form a
disk. Initially, the gaseous disk is only a few kpc in diameter but it
is slowly growing in size. After $1\gyr$ it has reached a diameter of
$20\kpc$. At $2\gyr$ the diameter is $\sim30\kpc$ which is comparable to
that of the stellar disk. At the same time the mass of the gaseous
disk is nearly constant. It has particle densities of $0.1 - 1
\cmmthree$ and temperatures (\reffig{fig_M4}) of less than
$10^4\Kelvin$. Towards the center the gas reaches higher temperatures
of up to several $10^6\Kelvin$.  Some diffuse gas remains in the halo
at densities of $\sim 10^{-4} \cmmthree$ and less while its
temperature is in the range of $10^6 - 10^7\Kelvin$. During the
initial collapse the diffuse gas gains a significant amount of angular
momentum transferred from the clouds by the drag force. For this
reason the gaseous disk rotates with a rotation velocity of
$\sim200\kms$.  Transient spiral patterns are also observed in the
gas disk.  Stellar feedback induces the emergence of bubbles of hot
gas.  These bubbles have diameters of up to $1\kpc$ and they cool
within a few $10\myr$. The volume filling factor of the hot gas
($T>10^4\Kelvin$) in the disk is 30--40\%.

The velocity dispersion of the stars increases with time. At $8\kpc$
the heating rate is about $1.3\kms\gyrmone$ and it is nearly constant
over the whole disk. In order to see how much heating is caused by
unphysical two-body relaxation (which stems from the inability to
resolve individual stars numerically) we compared the result with a
pure N-body simulation and found that the heating rate in our full
model is about 30\% higher than in the pure N-body calculation.  The
velocity dispersion of the clouds decreases at $0.6 \kms\gyrmone$ due
to the energy dissipation by cloud-cloud collisions. As a result
younger stars are born with a smaller velocity dispersion. This means
that a part of the observed 'heating' in the age-velocity-relation
might be indeed due to cooling as suggested by \citet{RFS04}.

The average SFR is $\sim 1.6\msun\yrmone$ (\reffig{fig_M6}).  Over the
$3\gyr$ of evolution the SFR decreases slowly from $\sim
4.0\msun\yrmone$ to $\sim 0.5\msun\yrmone$. A simple analytic model
indicates that this is due to the consumption of the clouds. The mean
SFE is roughly constant at a level of $\epsilon = 0.06$. The maximum
values of the local SFE are highest in the center with $\epsilon = 0.2
- 0.4$ (\reffig{fig_M7}). It can be seen in Fig.~\ref{fig_M8} that the
SFR also compares well to a recent determination of the Schmidt law by
\citet{K98}. Kennicutt has also shown that the SF law in disk galaxies
steepens abruptly below a threshold gas density of
$\sim10\msun\pcmtwo$. We also find in our simulations that the SF
  law steepens for lower gas densities. This is in agreement with 1d
  CD models of the settling of a gas disk which show a similar
  threshold for the formation of a thin disk \citep{BTH92}.
Excluding the data below $10\msun\pcmtwo$, we obtain a least-square
fit of our data to \refeq{eq_schmidtlaw} with $n \approx 1.7\pm0.1$
which is in fair agreement with $n \approx 1.4\pm0.15$ of \citet{K98}.

The mass exchange by condensation and evaporation 
reaches a quasi-equilibrium after $\sim 200\myr$. Before, evaporation
is the dominant process due to the heating of the diffuse gas induced
by the initial collapse of this component. In the later evolution, the
condensation rate is decreasing from $\sim 2\msun\yrmone$ to
$0.1\msun\yrmone$. The evaporation rate is slightly lower after
$1.5\gyr$ due to the decreasing SN rate, resulting in a reduced
heating of the diffuse ISM.

The total cloud mass is slowly decreasing,
showing the depletion by SF. The total mass of the diffuse gas remains
almost constant.  A detailed look shows that more than half of the
diffuse gas (65\%) is located in the disk ($R<20\,{\rm kpc}$ and
$z<1\,{\rm kpc}$), while the rest is in the halo.

In the initial model all clouds have the same mass. However, due to
fragmentation and coalescence a stationary cloud mass spectrum is
built-up within $1\gyr$ (\reffig{fig_M11}). This mass spectrum can be
fitted by a power-law with index $\alpha \approx -2$ with an
uncertainty of $0.2$. Once the mass spectrum is established it remains
unchanged until the end of the simulation.
%
%

%
%

\section{Discussion}
\label{sec_disc}

In the previous section we presented a model for a Milky-Way-type
galaxy. Since our description depends on several phase transitions and
exchange processes we want to discuss their influence in this
section. In detail we discuss the SF, cloud mass spectrum and
metallicities.

The model has a number of parameters which can be chosen within the
constraints of related observations or theory.  We therefore performed
simulations to investigate the influence of some of those parameters
using the model presented in \refsec{sec_refmod} as a reference. We
find that most models do not deviate strongly from the reference
model.

\subsection{Star Formation}
The mean SFR is $\sim 1.6\msun\yrmone$ in most models with a mean SF
efficiency of $\sim 6\%$. However, the SFR decreases in all models
from a few to below one solar mass per year. Observations on the other
hand reveal a much more constant SF history \citep{RSM00}. The
decrease of the SFR can be reproduced in an analytic model taking into
account the depletion of the cloudy medium by SF. An explanation for
the discrepancy might be that our simulation does not include any
infall to sustain a constant SFR. An infall rate of the order of
$\sim1\msun\yrmone$ would be sufficient for this provided the
infalling gas can be used in the SF process. The necessary infall rate
could be provided by high velocity clouds \citep{BST99,WHS99}.

\edit{removed paragraph}

The SF law is in good agreement with observations. Our data can be
fitted to \refeq{eq_schmidtlaw} with $n\approx1.7$ which is well
within the observed range of $n=1.4-2.0$ \citep{K98,WB02,BPB03}. It
should be noted that this result is widely independent of the choice
of our model parameters. Only models with a constant SF efficiency
show a different behaviour with $n=1.2$. \citet{KTH95} have shown that
stellar feedback leads to a self-regulated SF following a Schmidt-type
law. Our simulations confirm this result. We conclude that the
observed index $n$ of the SF law is a consequence of the
self-regulation process \new{which is enabled by allowing for a
variable SFE.}

\subsection{Cloud Mass Function}
\label{sec_msp}
The cloud mass function in our reference model has a slope of $\alpha
= -2$. In most observations the mass spectrum is shallower with $\alpha
\approx -1.5$ \citep[e.g.][]{SSS85} though there are a few observations
with larger values, e.g.\ $\alpha \approx -1.8$
\citep[e.g.][]{BW95}. \new{Taking into account the uncertainties in
both our model and observations we are able to resolve the cloudy
phase of the ISM, i.e.\ each cloud particle in our simulation
represents an individual molecular cloud.} 

Theoretical models of coagulation by \citet{FS65} also predict $\alpha
= -1.5$. These models include a fragmentation by SF but (different to
our model) SF occurs only when clouds reach a maximum mass. The
resulting fragments are then added to the lowest mass bin. In a \new{more}
similar model, which also takes into account a mass-radius relation
\refeqb{eq_mrrel}, \citet{E89} found a mass spectrum with $\alpha =
-2$, \new{showing that our simulations can reproduce these results.}

\subsection{Metallicity}
\label{ssec_zres}
We compared two models with metallicities $Z=1/2\ Z_\odot$ and
$Z=1.5\ Z_\odot$. The change of metallicity is only moderate and it
does not affect the dynamical evolution of the galaxy
significantly. This result verifies that using a
constant metallicity is a valid approximation on short time
scales. Using a simple estimate we find that for $3\gyr$ the expected
change in metallicity would be of the order $1/10\,Z_\odot$, i.e.\
well within in the range given by the two test
models. Therefore, a model including stellar yields would not change
the results as long as the late evolution (i.e.\ the last few Gyr) of
galaxies is simulated. However, stellar yields would have to be included
when the early formation stages are considered.

\subsection{The Multi-Phase ISM}
The ISM in our models can be divided into three phases: molecular
clouds and diffuse components, which can be either warm or hot. The
total gas mass fraction after $3\gyr$ of evolution is about $7\%$ in
all our models. The ratio of the mass of clouds to the mass of diffuse
gas is mainly determined by condensation and evaporation. If the
evaporation rate is higher, for example due to higher feedback energy
this ratio is closer to unity. $80-90\%$ of the diffuse gas is found
in the warm gaseous disk. Only $\sim15\%$ remain in the halo in a hot
gas phase. In the low-metallicity model this fraction is larger by a
factor of 2 due to less efficient cooling. \citet{SG03} have also
presented a model of a disk galaxy. In their model the ratio of cloud
and diffuse gas mass is $3$ which is comparable to our results. They
also find a hot gaseous halo.  In a similar calculation \citet{SC02}
find equal masses in clouds and gas independent of the feedback energy,
but their models do not include phase transitions between the gas
phases by condensation and evaporation. 

We have found a filling factor $f$ of $0.3-0.4$ for the hot gas in the
disk. \citet{MO77} predicted 0.7, a larger value, but our value is
in better agreement with the more recent findings of \citet{NI89} who
predicted a filling factor of about 0.2. Furthermore, three
dimensional simulation of a region of the galactic disk by \citet{A00}
show a filling factor for gas with $T>10^4\Kelvin$ of $\sim0.4$ in
good agreement with our results. We find an increase of $0.2$ in
$f$ if the energy injection by SNII is higher by a factor $4$ which is
comparable to the results of \citet{AB04}.

\new{One shortcoming in our treatment of the ISM is the lack of a 
detailed cooling below $10^4\Kelvin$. As a
consequence we do not allow clouds to form directly by thermal
instabilities. However, due to low mean densities and the turbulent
state of the diffuse gas we expect the formation of clouds in
expanding SN shells (which is included) to be much more important.
Again, a more sophisticated treatment of thermal instabilities would
be required for modelling the early evolution of (gas rich) galaxies.}

\subsection{Live DM Halo}

So far, in all presented models the DM halo was treated as a static
potential. For the case of the late evolution of an isolated galaxy a
static halo is reasonable.  However, DM particles have to be used,
e.g., for interacting galaxies or early galactic evolution. In one model
we therefore used a DM halo made of particles in order to test how
a live halo changes our results. The DM halo was modelled with
$10^5$ particles, thus DM particles are twice as massive as the
initial masses of star particles. As expected we do not find
significant changes in the evolution when using a live DM halo. The
CPU-time needed in this run is higher by a factor of 2 ($10$\,cpu-days
instead of $5$).

%
%

\section{Summary and Conclusions}
\label{sec_conclusions}

We have presented a new particle-based code for modelling the
evolution of galaxies. In addition to the collisionless stellar and
DM components it has incorporated a multi-phase description of
the ISM combining standard SPH with a sticky particle method to
describe diffuse warm/hot gas and cold molecular clouds respectively.
We have included a new description for SF with a local SF
efficiency depending on the local pressure of the ISM and the mass of
star forming clouds. Stellar life times and an IMF are taken into account
as well as SNII  and PNe. The gas phases can mix by means of
condensation and evaporation and clouds are subject to a drag force.

We used this description to model $3\gyr$ of evolution of a
Milky-Way-type galaxy. \edit{removed sentence}
\new{Overall,} the evolution of \new{our model} galaxy is quite
stable. Furthermore, a parameter study showed that the evolution does
not change significantly when varying model parameters like feedback
energy, cloud radius, SF time scale or metallicity in a reasonable
range.
\edit{removed sentence}

Our models produce a SF law with an index $n=1.7$ in good agreement
with observations. For models with a constant SF efficiency $\epsilon$
the index is with $n=1.2$ smaller than observations suggest. We
conclude, in agreement with \citet{KTH95}, that the SF law is the
result of \new{a} self-regulation \new{process enabled by our new
treatment of SF}. The average SFR is also comparable to
observations. Over time the SFR decreases by a factor of 4--8 because
we have not included any gas infall. To get a more constant SFR a gas
infall should be taken into account at a rate of
$\sim1\msun\yrmone$. The cloud mass spectrum in our models (with
$\alpha=-2$) is a little bit steeper than most observation
suggest. \new{Nonetheless, each cloud particle in our code essentially
represents an individual molecular cloud.}

\edit{removed sentence}

In all models most (85\%) of the diffuse gas forms a warm ($\sim
10^4\Kelvin$) rotating disk. A hot gaseous halo with temperatures of
$10^6-10^7\Kelvin$ and densities of $10^{-4}\cmmthree$ has been
built up. Hot gas is also found in the disk in bubbles accompanied by
SNII. The filling factor of the hot disk gas is about $0.4$ comparing
well to observations as well as models of the ISM. This result could
be used to find more realistic conditions for the initial model.

In a next step we will apply our model to interacting galaxies in order
to follow the evolution of the ISM and the SF history in perturbed
galactic systems. The chemical evolution should be embedded in order
to allow for a more detailed confrontation with observations.


\begin{acknowledgements}
      \new{The authors would like to thank P.~Berczik for fruitful
      discussions and the anonymous referee for his/her helpful
      comments to improve this manuscript.}  We thank W.~Dehnen
      (falcON), K.~Kujken \& J.~Dubinski (GalactICS) and P.~Teuben
      (NEMO) for making their software publicly available. This work
      was supported by the German \emph{Deut\-sche
      For\-schungs\-ge\-mein\-schaft, DFG\/} project number TH~511/2.
      S.~Harfst is grateful for financial support from the Australian
      Department of Education, Science and Training (DEST).
\end{acknowledgements}


\bibliographystyle{aa}           
\bibliography{harfst}


\appendix


\section{Determination of the star formation efficiency}
\label{app_starformation}

\citetalias{EE97} derived an
implicit relation between cloud mass \Mcld, pressure \Pism and SFE
$\epsilon$ (${\rm M}_5 = 10^5\msun$) 
\eq 1 = \epsilon +
\frac{1}{2}A_0\left(\frac{\Mcld}{{\rm M}_5}\right)^{-1/4}
\left(\frac{\Pism}{\Psun}\right)^{-5/8}\epsilon^2, {\rm when}\
\frac{\epsilon}{\xi}<1, \label{eq_eff1}
\qe

\noindent
and
\eq
\renewcommand{\arraystretch}{2.5}
\setlength{\arraycolsep}{0pt}
\begin{array}{ll}
    1 = \epsilon\ & +\
    A_0\left(\dfrac{\Mcld}{{\rm M}_5}\right)^{-1/4}
    \left(\dfrac{\Pism}{\Psun}\right)^{-5/8}\epsilon\xi
    \left[1+\dfrac{(\epsilon/\xi)^{2/5}-1.4}{0.56}\right]\\
    &+\ \dfrac{3}{14}
    A_0\left(\dfrac{\Mcld}{{\rm M}_5}\right)^{-1/4}
    \left(\dfrac{\Pism}{\Psun}\right)^{-5/8}\xi^2, \mathrm{when} \dfrac{\epsilon}{\xi}>1 \ .
\end{array}
    \label{eq_eff2}
\qe

\noindent
\citetalias{EE97} use $\Psun = \xtenton{3}{4}\cmmthree\Kelvin$ for the
pressure in the solar neighbourhood. $\xi$ is defined as
\eq
\xi \equiv \xi_0
\left(\frac{\Mcld}{{\rm M}_5}\right)^{-1/4}
\left(\frac{\Pism}{\Psun}\right)^{3/8}
\qe
\noindent
The constants $A_0 = 180$ and $\xi_0 = 0.1$ are determined from
observations of star forming clouds in the solar
neighbourhood. 

In each SF event \refeq{eq_eff1} and \refeq{eq_eff2} are solved
numerically to determine $\epsilon$. The resulting SFE is highest for
high mass clouds in a high pressure environment. Typical SFE are
between 5 and 10\%.


\section{SPH implementation}
\label{app_sph}

The idea of SPH is that a fluid is represented by $N$ particles.
Any physical property $A$ of the fluid at the position ${\bf r}_p$
is evaluated by means of a locally weighted average:
\eq
    A({\bf r}_p) = \sum_q m_q {A_q \over \rho_q} W({\bf r}_{pq},h).
\qe

\noindent
The summation is performed over particles $q$ with mass $m_q$, density
$\rho_q$. $A_q$ is the value of $A$ at ${\bf r}_q$ and the distance
${\bf r}_{pq}$ is ${\bf r}_p - {\bf r}_q$. We apply the spherically
symmetric smoothing kernel $W(r,h)$ suggested by \citet{ML85}
\eq
   W(r,h) = {1 \over \pi h^3}
            \left\{
            \begin{array}{lllll}
             \displaystyle
            1-{3\over2}\left({r\over h}\right)^2
             +{3\over4}\left({r\over h}\right)^3,
                           & \mbox{ for $0  \leq {r\over h} \leq 1$} \\
                                                                        \\
             \displaystyle
            {1\over4}\left(2 - {r\over h}\right)^3,
                           & \mbox{ for $1  \leq {r\over h} \leq 2$} \\
                                                                        \\
            0,
                           & \mbox{ for ${r\over h} > 2$.} \\
            \end{array}
            \right.
\qe

\noindent
A variable smoothing length $h$ keeping the number $N_{\rm sm}$ of nearest
neighbours nearly constant at about 32 for all SPH particles is applied.

The equation of motion of an SPH particle $p$ is given by
\eq
     {d^2{\bf r}_p \over dt^2} = - {1 \over \rho_p} \nabla P_p
                                 + {\bf a}_p^{\rm visc} - \nabla \Phi_p,
\qe

\noindent
where $\Phi_p$ is the gravitational potential, $P_p$ the pressure and
${\bf a}_p^{\rm visc}$ an artificial viscosity term to handle
shocks. We use a symmetric expression to estimate $(\nabla P)/\rho$:
\eq
  {\nabla P_ p \over \rho_p} = \sum_q m_q \left( {P_p \over \rho_p^2}
                               + {P_q \over \rho_q^2}\right) \nabla W_{pq},
\qe

\noindent
where $W_{pq}$ indicates that the arithmetic mean of the two kernel
function for the individual smoothing lengths $h_p$ and $h_q$ is used.
For the artificial viscosity term ${\bf a}_p^{\rm visc}$ we use
\citep{GM83}
\eq
   {\bf a}_p^{\rm visc} = - \sum_q m_q \Pi_{pq}  \nabla W_{pq},
\qe

\noindent
with
\eq
  \Pi_{pq}  = { -\alpha_{\rm v}\mu_{pq}c_{pq} + \beta_{\rm v}\mu_{pq}^2\over \rho_{pq}},
\qe

\noindent
where $\rho_{pq}$ and $c_{pq}$ are the arithmetic means of density and
sound speed of particles $p$ and $q$. $\mu_{pq}$ is defined by
\eqa
    \mu_{pq} = \left\{ \begin{array}{ll}
                   {\displaystyle{\bf v}_{pq}\cdot{\bf r}_{pq} \over
                     \displaystyle h_{pq}\left( r_{pq}^2 / h_{pq}^2 + \eta_{\rm v}^2 \right)}
                     & \mbox{for ${\bf v}_{pq}\cdot{\bf r}_{pq} < 0$}    \\
                                                                         \\
                   0 & \mbox{for ${\bf v}_{pq}\cdot{\bf r}_{pq} \geq 0$} \\
                      \end{array}
               \right.,
\aqe

\noindent
where ${\bf v}_{pq}$ is the relative velocity of particle $p$ and $q$.
The viscosity parameters are set to $\alpha_{\rm v}=2$, $\beta_{\rm v}
= 1$ and $\eta_{\rm v}=0.1$. The evolution of the specific internal
energy $u$ is computed by
\eq
    {du_p \over dt} = {1\over2}\sum_q m_q \left({P_p \over \rho_p^2}
                               + {P_q \over \rho_q^2} + \Pi_{pq} \right)
                         {\bf v}_{pq}\cdot\nabla W_{pq}
                       + {\Gamma - \Lambda \over \rho}.
\label{eq_energy}
\qe

\noindent
$\Gamma$ and $\Lambda$ account for heating and cooling
processes. 

The time integration is done by a leap-frog integrator.
Individual time steps are used for each particle. The time
step for an SPH particle $p$ is chosen to satisfy a modified form of
the Courant condition as suggested by \citet{M92}
\eq
    \Delta t_p \le C { h_{p} \over h_{p}\left|\nabla\cdot{\bf v}_p\right|
                                   + c_p + 1.2 \left(\alpha_{\rm v} c_p
                                   + \beta_{\rm v} \max_q \left| \mu_{pq}\right| \right)},
\qe

\noindent
where $C = 0.4$ is the Courant number, $h_{p}$ the smoothing
length, $c_p$ the sound speed, ${\bf v}_p$ the velocity of particle $p$.

Because cooling time scales are usually much shorter than the
dynamical times, impractically short time steps might occur.
Therefore, \refeq{eq_energy} is often solved semi-implicitly. 
However, we chose a
different approach: a fourth-order Runge-Kutta scheme is used to
integrate the cooling function beforehand with a sufficiently small
time step and a look-up table with times and temperatures is
created. The number density $\ngas$ is fixed to one value for this but
cooling times scale linear with $\ngas$ and can be easily
rescaled. Starting from a given temperature the new temperature after
a (rescaled) time step is then simply looked up from the table. This
provides a fast way of computing the change in temperature, i.e.\ the
cooling rate, for any given temperature, density and time step. The
cooling rate $\Lambda$ for each particle is updated each individual
time step. 

A more detailed description of the SPH method can be found
in \citet{M92}.

\end{document}